# SELF-ORGANIZATION OF A 2D LATTICE ON A SURFACE OF GE SINGLE CRYSTAL AFTER IRRADIATION WITH YAG: ND LASER

*A. Medvid', A. Mychko, P. Onufrievs*

Riga Technical University, Laboratory of Semiconductor Physics, LV-1048, 14 Azenes Str., Latvia

Abstract

Experimentally observed self-organization of a 2D lattice on the surface of Ge single crystal after irradiation by pulsed YAG:Nd laser is reported. The 2D lattice consists of nano-size elevations arranged in a pattern of $C_{6i}$ point group symmetry and is characterized by translational symmetry with the period of 1μm. Calculations of time-dependent distribution of temperature in the bulk of the Ge sample are presented to explain the phenomenon. The calculations show that overheating of the crystal lattice occurs at laser radiation intensities exceeding 30 MW/cm$^2$. According to synergetic ideas, the presence of the non-equilibrium liquid phase of Ge and huge gradient of temperature ($\sim 3 \times 10^8$ K/m) can lead to self-organization of the 2D lattice similar to Benard cells.

## "1. INTRODUCTION"

Self-organized structures (SOSs) occurring in nature are often objects of studies in physics, chemistry, and biology on both macroscopic and microscopic levels. They arise at phase transitions of a system from an unstable state to a stable state. The new stable state is characterized by emergence of regular structures appearing spontaneously without any outside influence on the system. SOSs processes are investigated by synergetics [1] describing the processes mainly mathematically without consideration of the microscopic nature of the phenomena. SOSs often arise in processes of microelectronics and optoelectronics technologies [2], when systems are strongly excited. This process happens at heteroepitaxial growth of Ge quantum dots (QD) on the surface of Si [3] or in a Ge/Si system [4, 5] due to mechanical stress caused by mismatch between crystal lattices of Ge and Si. Self-organized nano-structures can be produced by oxygen plasma sputtering [6]. The random localization of nano-structures is uncontrollable. Powerful laser radiation (LR) very often induces periodic micro-SOSs on surfaces of semiconductors and metals - the so-called laser-induced periodic surface structures (LIPSS) [7]. Moreover, LR is used for ablation of Si by eximer laser for the formation of nano-peaks producing bright photo-luminescence in the visible spectrum [8]. Due to insufficient knowledge of LR interaction with semiconductors, laser technologies are poorly controllable and, as a result, are seldom used in microelectronics technology. Therefore, study of interaction between LR and semiconductor is a very important task for semiconductor physics and microelectronics technology, especially at high levels of excitation, when irreversible changes of optical and electrical properties of a semiconductor take place. At the first stage of this interaction the energy of LR is transformed into heat, but further stages of these processes are not determinates [9]. Moreover, the morphology of the irradiated surface of a semiconductor depends on conditions of irradiation (for example, a scanning laser beam or pulse-by-pulse irradiation of the same spot [8]) and homogeneity of laser beam; single mode or multimode irradiation regime.

In this paper, we present the results of the study of interaction between one mode YAG:Nd laser radiation and Ge single crystal at intensities leading to self-organization of 2D lattices of $C_{6i}$ point group in a symmetric pattern.

## "2. EXPERIMENT AND DISCUSSION"

The experiments were performed in ambient atmosphere at pressure of 1 atm, T = 20 C, and 60% humidity. Radiation from a pulsed YAG:Nd laser (15 ns pulses of $\lambda = 1.06$ μm, pulse rate 12,5Hz, intensity I=30MW/cm$^2$) was directed normally to the surface of i-type single crystal Ge(1 1 1) or Ge(0 0 1) samples with sizes 1.0 x 0.5 x 0.5 cm$^3$ and resistance r = 45 Ω cm. The samples were polished mechanically and etched in CP-4A solution to ensure the minimum surface recombination velocity $S_{min}=10^2$cm/s on all the surfaces. The spot of laser beam of 3mm diameter was scanned along the sample surface by two-coordinate manipulator in 1 μm step.

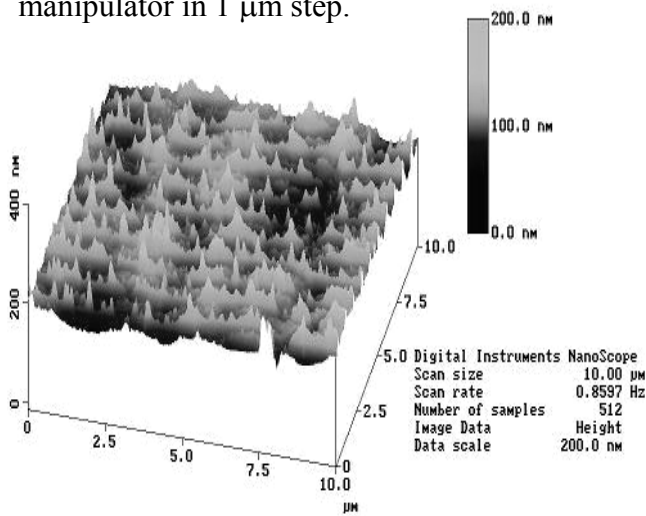

Fig.1. Three-dimensional image of a monocrystal surface of Ge irradiated by laser with intensity 30 MW/cm$^2$

In the range of intensities from 25MW/cm$^2$ to 30MW/cm$^2$, nanostructure is created on the semiconductor surface. The creation of this structure is connected to substantial movement of material from bottom upwards.

Three-dimensional image of a monocrystal surface of Ge irradiated with laser radiation of intensity 30 MW/cm$^2$ is shown on fig.1.

In fig.1 we can see that the surface micro-relief contains not only two-dimensional periodic structure (LIPSS), but also nano-hills, situated on top of two-dimensional periodic structures where hexagonal structure of point group $C_{6i}$ with translation symmetry is covering all the surface of Ge sample, irradiated by laser (fig.2.). The repetition period of this structure is 1μm. At a qualitative level it is possible to explain occurrence of the cells, with the synergetic effect, connected with self-organizing in cellular structures, such as Benard cells in the initial stage of their organization.

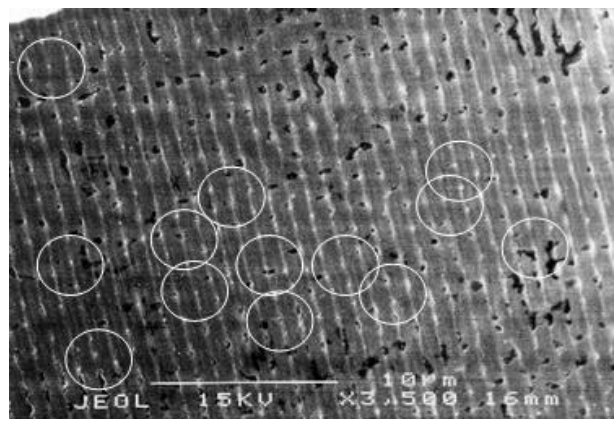

Fig. 2 Image of Ge surface irradiated with the laser intensity about 30MW/cm$^2$ received with the help of a scanning electronic microscope [10].

In compliance with synergetic idea: at present of non equilibrium liquid faze and high temperature gradient 3·10$^8$ K/cm a turbulent stream is created and advanced to create the hexagonal structures from theoretical point of view. This hexagonal structure of cells is the beginning stage of Benard cells formation. The cells size is comparable with thickness of liquid [10].

During laser pulse, the surface material is overheated by $400^0$C and temperature gradient achieves $3 \cdot 10^8$ K/cm. The small outside action in our case is a light pressure, sufficient for transferring a metastable system to unstable state. As the result, the movement of liquid phase is started. Two streams are created. The first one is directed into the bulk of the sample towards decreasing temperature due to the pressure of laser irradiation, and the second – towards the radiated surface where decreasing of liquid density. According to the law of non interrupted flow of liquid, two flows are created in the opposite directions. The turbulent flow that appears in our case is initiates formation of Benard convection cell in the initial stage.

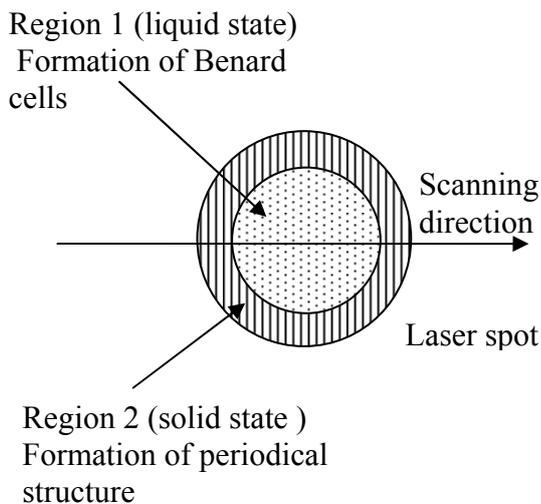

Fig.3. The position of laser spot on the surface sample in scanning mode.

Formation of 2D lattice, periodical and hexagonal in the shape of Benard cells, can be explained using scanning regime and distribution of light intensity in laser beam by Gauss law. There are two temperature gradients in this case, the first in the bulk of the sample, the second along the radius of laser beam spot. Formation of the periodical structure is the result of multiple laser pulse irradiation in region of 2 (fig.3.) from behind of temperature gradient action along to the surface of the sample. The movement of laser spot along the surface of the sample with speed of scanning 14μm/s and laser pulse repetition rate 12 Hz is stimulated until creation of periodical structure with period of 1μm (fig.1). The process of creation of this structure is described by theory of laser-induced defect-deformational instabilities on surfaces [11]. The irradiation of the surface of the sample with periodical structure by intensity (the centre of Gauss pulse, region 1.) which initiates material overheating and melting under action of huge temperature gradient directed in the bulk of the sample, leads to creation of Benard cells. The further movement of laser spot along the surface of the sample with Benard cells lead to masking of cells by creation of periodical structure.

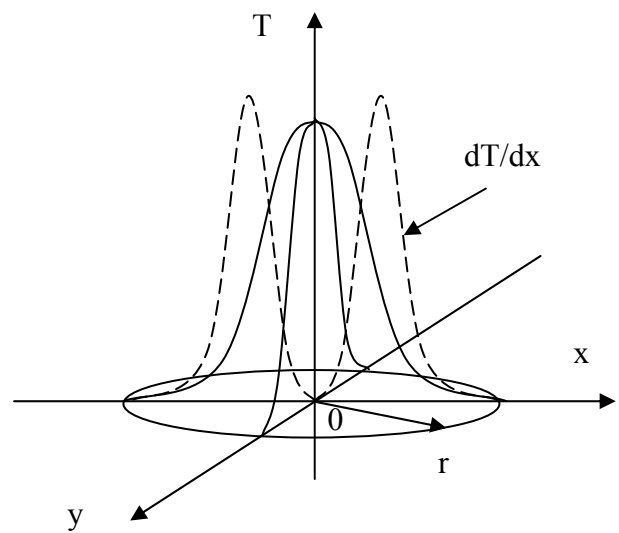

Fig.4. The distributions of temperature (solid line) and temperature gradient (dashed line) after laser irradiation: dT/dx - along the radius of laser beam spot.

## "3. CONCLUSION"

A 2D lattice consisting of nano-scale elevation patterns of $C_{6i}$ point group symmetry is formed on the surface of Ge crystals under YAG:Nd laser radiation at intensities 30 MW/cm$^2$.

## ACKNOWLEDGEMENTS

The study has been supported by the Latvian Council of Science Grant No. 01.0577.

## REFERENCES

[1] H. Haken, Advanced Synergetics, Spring-Verlag, Berlin, 1983.
[2] A.G. Nassiopoulan, S. Nazaki, in: Berbezier (Ed.), Micro- and Nanostructured Semiconductors, Elsevier, Amsterdam, 2003.
[3] N. Motta, A. Sgarlata, F. Risei, P.D. Szkutnik, S. Nufris. M. Scarselli. A. Balzarotti, Controlling the quantum dot nucleation site, Mater. Sci. Eng. B, 101, pp.77-88, 2003.
[4] A. Ronda, I. Berzbezier, A. Pascale, A. Portavoce. F. Volpi, Experimental insights into Si and SiGe growth instabilities: Influence of kinetic growth parameters and substrate orientation, Mater. Sci. Eng. B, 101, pp.95-101, 2003.
[5] M. Goryll, L. Vescan, H. Luth, Nucleation of Ge islands on Si mesas with high-index facets, Mater. Sci. Eng. B, 101, pp.9-13, 2003.
[6] Y. Sun, L. Miyasata, J.K. Wigmore, Self-organized growth of zero-, one-, and two-dimensional nanoscale SiC structures by oxygen-enhanced hydrogen plasma sputtering, J. Appl. Phys. 86, pp. 3076-3082,1999.
[7] F. Jeff, J.E. Sipe, H.M. van Driel, Laser-induced periodic surface structure. III. Fluence regimes, the role of feedback, and details of the induced topography in germanium, Phys. Rev. B 30, pp. 2001-2015,1984.
[8] A.V. Kabashin, M. Meumier, Laser-induced treatment of silicon in air and formation of Si/SiO$_x$ photoluminescent nanostructured layers, Mater. Sci. Eng. B, 101, pp. 60-64, 2003.
[9] D.K. Biegelsen, G.A. Rozgonyi, C.V. Shank, Energy Beam-Solid Interactions and Transient Thermal Processing, vol. 35, Mater. Res. Soc, Pittsburgh, 1985.
[10] A. Medvid', Y. Fukuda, A. Michko, P. Onufrievs, Y. Anma, 2D lattice formation by YAG:Nd laser on the surface of Ge single crystal, Appl. Surf. Sci., Vol 244/1-4, pp.120-123, 2005.
[11] V.I.Emel'yanov, A.A.Soumbatov, Periodic Phase Structure Formation in Pulse Induced Crystallization of Films, Phys.Stat.Sol.a, vol.158, pp. 493-497,1996.